\documentclass[sigconf]{acmart}

\copyrightyear{2026}
\acmYear{2026}
\setcopyright{cc}
\setcctype{by}
\acmConference[EASE '26]{International Conference on Evaluation and Assessment in Software Engineering}{June 09--12, 2026}{Glasgow, United Kingdom}
\acmBooktitle{International Conference on Evaluation and Assessment in Software Engineering (EASE '26), June 09--12, 2026, Glasgow, United Kingdom}
\acmDOI{10.1145/3816483.3816485}
\acmISBN{979-8-4007-2348-3/2026/06}

\usepackage{graphicx} 

\title[Do Privacy Policies Match with the Logs? An Empirical Study of Privacy Disclosure]{Do Privacy Policies Match with the Logs? An Empirical Study of Privacy Disclosure in Android Application Logs}

\usepackage{booktabs}
\usepackage{amsmath}
\usepackage{amsmath}
\usepackage{hyperref}

\DeclareRobustCommand{\inline}[1]{%
  \texttt{\small\detokenize{#1}}%
}

\usepackage{subfig}

\usepackage{cleveref}

\usepackage{xcolor}

\definecolor{customgray}{cmyk}{0, 0, 0, 0.7}

\newsavebox{\summarybox}

\newenvironment{Summary}[2]{%
  \par\smallskip
  \setlength{\fboxrule}{1.2pt}%
  \setlength{\fboxsep}{6pt}%
  \begin{lrbox}{\summarybox}%
  \begin{minipage}{\dimexpr\linewidth-2\fboxsep-2\fboxrule\relax}
  \colorbox{customgray}{%
    \makebox[\dimexpr\linewidth-2\fboxsep\relax][l]{%
      \color{white}\bfseries\small #1%
    }%
  }%
  \par\smallskip
}{%
  \end{minipage}%
  \end{lrbox}%
  \noindent\fcolorbox{customgray}{white}{\usebox{\summarybox}}%
  \par\smallskip
}

\begin{document}

\author{Zhiyuan Chen}
\affiliation{
  \institution{Rochester Institute of Technology}
  \city{Rochester}
  \state{NY}
  \country{USA}
}
\email{zc9482@rit.edu}

\author{Love Jayesh Ahir}
\affiliation{
  \institution{Rochester Institute of Technology}
  \city{Rochester}
  \state{NY}
  \country{USA}
}
\email{la3679@rit.edu}

\author{Ahmad Suleiman}
\affiliation{
  \institution{Rochester Institute of Technology}
  \city{Rochester}
  \state{NY}
  \country{USA}
}
\email{as4300@rit.edu}

\author{Kundi Yao}
\affiliation{
  \institution{Ontario Tech University}
  \city{Oshawa}
  \state{ON}
  \country{Canada}
}
\email{Kundi.Yao@ontariotechu.ca}

\author{Yiming Tang}
\affiliation{
  \institution{Rochester Institute of Technology}
  \city{Rochester}
  \state{NY}
  \country{USA}
}
\email{yxtvse@rit.edu}

\author{Weiyi Shang}
\affiliation{
  \institution{University of Waterloo}
  \city{Waterloo}
  \state{ON}
  \country{Canada}
}
\email{wshang@uwaterloo.ca}

\author{Daqing Hou}
\affiliation{
  \institution{Rochester Institute of Technology}
  \city{Rochester}
  \state{NY}
  \country{USA}
}
\email{dqvse@rit.edu}

\begin{abstract}
Privacy policies are intended to inform users about how software systems collect and handle data, yet they often remain vague or incomplete. 
This paper presents an empirical study of patterns in log-related statements within privacy policies and their alignment with privacy disclosures observed in Android application logs.
We analyzed 1,000 Android apps across multiple categories, generating 86,836,964 log entries. 
Our findings reveal that while most applications (88.0\%) provide privacy policies, only 28.5\% explicitly mention logging practices. Among those that reference logging, most clearly describe what information is logged; however, 27.7\% of log-related statements remain overly simplistic or vague, offering limited insight into actual data collection. We further observed widespread privacy leakages in application logs, with 67.6\% of apps leaking sensitive information not mentioned in their policies. Alarmingly, only 0.4\% of applications demonstrated consistent alignment between declared policy contents and actual logged data.
These findings highlight that current privacy policies provide incomplete or ambiguous descriptions of logging practices, which frequently do not align with actual logging behaviors.

\end{abstract}


\maketitle

\section{Introduction}

Logs are essential resources of modern software systems, providing valuable information for software development activities, including debugging, performance monitoring, and security auditing. They record runtime events, errors, and user interactions, helping developers understand how applications behave in real-world usage scenarios~\cite{Zishuo2023,Wenhao2017,Chen2017}.
For mobile applications in particular, 
the importance of logs is also undeniable due to their diverse runtime contexts and limited opportunities for direct observation.

Despite their technical value, logs often contain sensitive information that raises privacy concerns.
Application logs frequently capture information including device identifiers, IP addresses, location data, user credentials, and user behavior patterns that could inadvertently compromise user privacy if improperly handled~\cite{Christof2016, Zhiyuan2024Vulnerability, Stephan2019}. 
This risk is particularly acute in Android ecosystem, which powers over 3 billion active devices worldwide and represents approximately 70\% of the global smartphone market~\cite{Josh2024}. 
As users interact with applications, logs are continuously generated and can inadvertently expose personal or contextual information beyond the user’s awareness. From the user’s perspective, the impact of such privacy leakage is often unpredictable and rarely confined to a limited scope as many applications transmit logs to remote servers or third-party analytics platforms and users’ private data can be disseminated far beyond their devices.
Despite this, the collection and handling of application log data are rarely made transparent to users. Most users rely on privacy policies to understand how their data are collected, used, and shared. Privacy policies are designed to inform users and ensure compliance with regulations such as the General Data Protection Regulation (GDPR)~\cite{GDPR} and the California Consumer Privacy Act (CCPA)~\cite{CCPA}. 
This lack of explicit disclosure creates a potential gap between what applications claim to do and what they actually log in practice.

This gap between logging practices and their disclosure in privacy policies raises a fundamental question: \textit{What are the prevalence and nature of log-related statements in Android app privacy policies, and do these statements align with actual logging behaviors observed in application logs?} 
To address this question, we conduct a large-scale empirical study to investigate the relationship between Android application logs and their corresponding privacy policies. 
We analyze 1,000 Android applications across 42 categories, generating and examining 86,836,964 log entries to identify patterns in the disclosure of logging practices within privacy policies and their alignment with real-world logging behaviors.
In particular, we focus on three research questions (RQs):
\begin{itemize}
    \item \textbf{RQ1:} How prevalent is the disclosure of logging practices in Android applications' privacy policies, and what factors influence the presence of such disclosures?
    \item \textbf{RQ2:} How do privacy policies specify the logging practices of Android applications?
    \item \textbf{RQ3:} To what extent do log-related statements in privacy policies align with the privacy disclosures present in real-world Android application logs?
\end{itemize}

Our findings reveal substantial discrepancies in transparency and disclosure practices across the Android ecosystem. While most applications (88.0\%) provide privacy policies, only 28.5\% explicitly mention logging practices. Applications with higher download volumes and more user reviews are more likely to disclose logging practices.
Among the policies that reference logging, most clearly describe what information is logged; however, 27.7\% of log-related statements remain overly simplistic or vague, offering limited insight into actual data collection activities.
We identified alarming inconsistencies between stated policies and actual logging behaviors, where 67.6\% of analyzed applications exhibited privacy leakages in their logs, with 76.92\% leaking information cases not mentioned in their privacy policies. Particularly concerning is that only 0.4\% of applications demonstrated consistent alignment between their privacy policies and actual logged data. 
These findings highlight an urgent need for improved transparency mechanisms and regulatory oversight to bridge the significant gap between privacy promises and actual data collection practices in the mobile application ecosystem.

Our contributions include:
\begin{itemize}
    \item A comprehensive dataset consisting of 1,000 Android applications, including their privacy policies and runtime logs\footnote{The replication package is available at: \url{https://github.com/SEDA-RIT/PrivacyPolicy} }.
    \item Empirical insights into the factors that influence the presence of log disclosure in mobile application privacy policies.
    \item An in-depth analysis of how privacy policies disclose logging activities — specifically, what is logged, why logs are collected, and how these practices are described.
    \item A novel privacy leakage detection method combining keyword-based detection with LLM-assisted keyword expansion to identify sensitive information in Android app logs.
\end{itemize}


\noindent \textit{Paper organization. }
The remaining sections of this paper are organized as follows. Section \ref{sec:background} introduces the background and reviews related work. Section \ref{sec:overview} describes the study design and experimental setup. Section \ref{sec:results} presents the experimental results, addressing the three RQs and summarizing the main findings. Section \ref{sec:discussion} provides further discussion and interpretation of the results.
Section \ref{sec:threat} examines potential threats to validity. Finally, Section \ref{sec:conc} concludes the paper and outlines directions for future work.

\section{Background and Related Work}
\label{sec:background}
\subsection{The Prevalence and Quality of Privacy Policy Disclosures}

Privacy policies play a central role in informing users about how mobile applications collect, use, and share personal data, while also serving as a key mechanism for regulatory compliance~\cite{ZimmeckWZILSWSB16,ZimmeckSSRWRRS19}, such as the General Data Protection Regulation (GDPR) and the California Consumer Privacy Act (CCPA).
Despite their intended purpose, prior studies have shown that many applications either omit privacy policies entirely or provide only partial disclosures of their data practices~\cite{StoryZS18}. Even when policies are available, they often fail to specify what data are collected or how such data are processed and shared~\cite{liu2022evaluating}. Many privacy policies also contain ambiguous descriptions, and are difficult for users to comprehend~\cite{ZimmeckWZILSWSB16,WilsonSDLCLAZSR16,YuLLZ16,Jaspreet2015}. The prevalence and quality of disclosures have been found to vary significantly across app categories, developer regions, and popularity levels~\cite{ZimmeckSSRWRRS19,StoryZS18}. 
For example, popular categories such as \textit{Books\_and\_Reference}, \textit{Education}, and \textit{Entertainment} show some of the lowest privacy-policy coverage among all app types~\cite{StoryZS18}.
These results show that many privacy policies remain inconsistent and opaque, making it difficult for users to trust how applications handle their personal data. As an essential data source in Android systems, logs capture a wide range of runtime information and user interactions~\cite{Ding2024, Ding2023, Zhang2022, He2016}. The responsible and transparent handling of such logging data should therefore be clearly addressed in privacy policies. 
Such logging data frequently includes personal data and identifiers~\cite{Zhiyuan2024Vulnerability}, placing it within the regulatory scope of data protection laws that require transparency, purpose limitation, and data minimization. The responsible and transparent handling of logging data should therefore be explicitly addressed in privacy policies.
However, to the best of our knowledge, no prior work has systematically examined how logging practices are disclosed in Android applications’ privacy policies. This gap motivates a closer investigation into how the collection and use of log data are specified and communicated within these documents.


\subsection{Specification of Data Collection in Privacy Policies}


Recent studies have extracted data collection statements from privacy policies. These approaches aim to identify data types, collection purposes, and third-party sharing practices at scale, improving the transparency of policy analysis~\cite{zimmeck2019maps,ravichander2019question,andow2019policylint}.
Yousra et al.~\cite{Yousra2024} reviewed privacy policies by examining how organizations describe their data collection and usage practices, including what personal information is gathered, how it is collected, and for what purposes it is used.
Wagner et al.~\cite{wagner2022privacypoliciesagescontent} compiled a large-scale corpus of privacy policies (1996–2021) and analyzed their length, readability, and data practice descriptions using machine learning and natural language processing.  
Prior work has also shown that users rarely read privacy policies due to their excessive length, technical language, and average reading time of approximately 12 minutes~\cite{mcdonald2008cost,obar2020biggest}.



\subsection{Policy–Behavior Alignment}



Many studies have emphasized the lack of completeness, precision, and consistency in privacy policy disclosures. 
Empirical analyses show that privacy policies are frequently missing, incomplete, or misaligned with actual data practices. 
For example, a review of popular health applications found that only 183 out of 600 provided any privacy policy at all~\cite{Chang2020}. 
Similarly, Zhan et al.~\cite{Yuxia2024} reported a general absence of privacy policies in virtual reality (VR) applications, 
while Srinath et al.~\cite{Mukund2023} identified numerous instances of dead links, empty pages, and placeholder text in web application policies. 
Extending beyond policy availability, Bui et al.~\cite{bui2021consistency} introduced \textit{PurPliance}, which examined whether applications’ stated purposes for data use align with their actual behaviors, revealing frequent mismatches between declared intentions and runtime practices.


Several works have evaluated whether app privacy policies fully enumerate the data they collect.
Rahman et al.~\cite{baalous2025detecting} developed an automated pipeline, \textit{PermPress}, to verify whether Android apps’ privacy policies mention all the sensitive permissions they actually use. 
Building on this line of work, Andow et al.~\cite{Benjamin2020} introduced \textit{PoliCheck} to analyze privacy-sensitive data flows in 13,796 Android apps and their corresponding policies, revealing that up to 42.4\% of apps misrepresented or completely omitted certain data flows in their privacy disclosures.

Another thread of research looks at how third-party code in apps can lead to policy non-compliance. Tan and Song~\cite{khedkar2024advancing} developed a tool called PTPDroid to detect violations in user privacy disclosures specifically related to third-party libraries in Android apps.
Overall, such alignment studies consistently show that what apps do often does not fully match what their privacy policies say, underscoring the need for better enforcement and more truthful privacy disclosures~\cite{Benjamin2020}.

\section{Empirical Study Overview}
\label{sec:overview}

\subsection{Methodology Overview}

Our research design consists of three research questions (RQ1–RQ3), each addressing a different aspect of how Android applications handle and disclose logging practices in their privacy policies.
This multi-stage design allows us to move from disclosure occurrence (RQ1) to content characterization (RQ2) and finally to behavioral validation (RQ3), providing both breadth and depth in understanding the gap between declared and actual logging practices.

\begin{figure*}[h]
    \centering
    \includegraphics[width=0.85\textwidth]{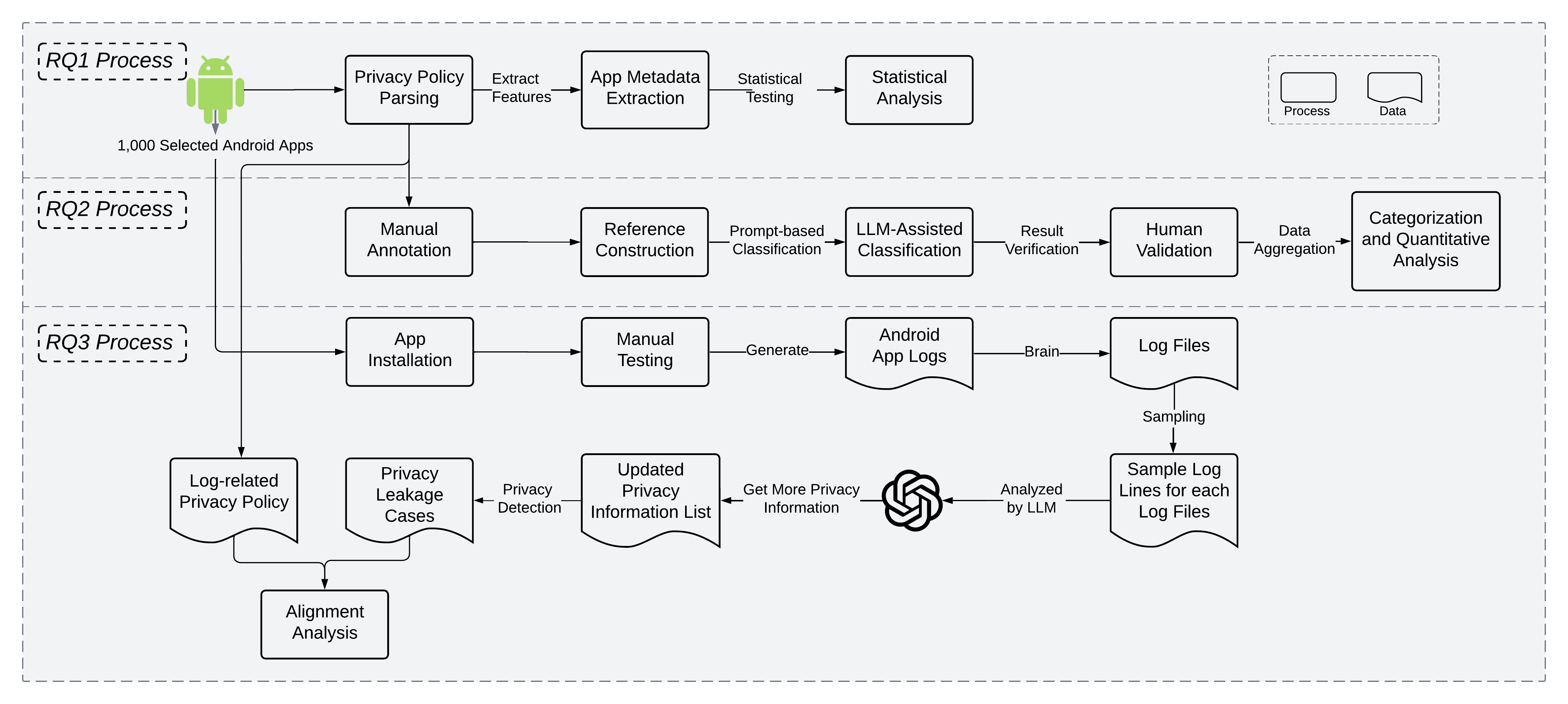}
	\caption{An overview of our study.} 
	\label{fig:workflow} 
    \vspace{-1em}
 \end{figure*}

Figure \ref{fig:workflow} presents the overall workflow of our study.
For RQ1, we analyzed the prevalence of log-related disclosures in privacy policies and factors influencing their presence. We selected 1,000 apps from Google Play, parsed their privacy policies for log-related terms, and manually verified the matches. App metadata such as category, downloads, ratings, and reviews were extracted, and statistical tests (Chi-square and Welch’s t-tests) were conducted to identify influencing factors.
For RQ2, we examined how applications specify logging activities in their privacy policies. 
We adopt a semi-automated approach, in which log-related descriptions extracted from privacy policies are preprocessed using an LLM and then manually verified to determine what is logged, why logs are collected, and how these practices are described in the policies.
For RQ3, we investigated the alignment between declared policies and actual logging behaviors. 
We installed 1,000 applications and conducted manual testing to generate runtime logs. We then performed an alignment analysis by comparing the detected leakage cases with the corresponding privacy policy statements to identify inconsistencies. 

\subsection{Empirical Setup}

\subsubsection*{\textbf{App Selection}}




The selection of 1,000 Android applications was carefully curated to ensure a diverse and representative dataset for empirical analysis. 
We prioritized applications from popular categories such as Social Media, Utilities, Communication, Productivity, and Entertainment when selecting apps for our study.
We first selected a set of popular applications as seed apps based on Google Play Store rankings across a diverse set of categories. We then identified additional candidate applications through the recommendation and related-app features provided by the app store interface.
Each candidate app was manually reviewed using its Google Play preview screenshots and textual descriptions. An app was selected if it provided at least one user-facing interactive feature that enabled active user input or contribution, such as buttons, chat boxes, task creation panels, and message interfaces. Consequently, applications that require active user engagement are more likely to produce rich behavioral data and generate extensive log content.
We conducted our study on a dataset of 1,000 Android applications selected from the Google Play Store. The chosen apps span across 42 categories, covering diverse domains such as social, finance, health, tools, and games.
This multi-stage selection strategy could balance categorical diversity, popularity, and demonstrable interactivity. It ensured the resulting dataset was both comprehensive and representative of real-world Android application behaviors.

\subsubsection*{\textbf{Log Generation and Collection}}





For each selected application, the data collection process began by launching the app and preparing the testing environment on a Google Pixel 5 device.
Prior to each session, all existing system logs were cleared using the command \inline{adb logcat -c} to ensure that subsequent logs reflected only new runtime activities.
Each application was then manually exercised for a fixed duration of 15 minutes, during which the tester interacted with a variety of features to emulate realistic user behavior according to specific application components. 
All the applications were tested by the same tester.
The interaction strategy was adapted to each application, with the goal of testing all interactions that could generate more runtime log records. All applications were tested by a single tester.
Throughout the session, logs were continuously captured in the background using the command \inline{adb logcat > <app_package>_<timestamp>.log}.
At the end of each session, logging was terminated, and the output file was saved under a unique filename containing both the app’s package name and a timestamp.
This procedure ensured that every log file could be accurately associated with its corresponding application and testing session, thereby maintaining traceability and consistency across the dataset.




The experiment was carried out over a period of 11 months, starting from November 2024 to September 2025.
As a result, we generated 1,000 log files, one for each application. These logs contained 86,836,964 lines of log entries. The number of lines per application varied substantially, with the smallest log file containing 11,879 lines and the largest containing 1,265,699 lines, while the average size was 86,837 lines per log file. This distribution reflects the diversity of logging behaviors across applications.

\section{Results} \label{sec:results}

In this section, we present the results of the study in terms of RQ1-3.

\subsection*{\textbf{RQ1: How prevalent is the disclosure of logging practices in Android applications’ privacy policies, and what factors influence the presence of such disclosures?}}




\textbf{\textit{Motivation.}}
Privacy policies are designed to be the principal channel of communication between application developers and end-users regarding data collection and usage. However, prior research consistently shows that these policies are often vague, incomplete, or misaligned with an application's actual behavior~\cite{ZimmeckWZILSWSB16,WilsonSDLCLAZSR16,YuLLZ16,Jaspreet2015}.
In Android applications, logging serves as a key mechanism through which developers obtain runtime information, often the only way to monitor performance and diagnose issues. Because logs may capture sensitive or identifiable data, it is essential that their collection and usage should be clearly communicated to users.
Therefore, understanding how frequently privacy policies mention logging, and identifying the factors that drive such disclosures, is essential for evaluating the adequacy of privacy communication in Android applications.

\textbf{\textit{Approach.}}
To examine which factors influence whether Android applications disclose logging practices in their privacy policies, we performed a statistical analysis combining text mining and inferential statistics.
For each application, we collected its privacy policy by accessing its Google Play Store page, following the developer-provided privacy policy link, and extracting the policy content from the corresponding external webpage.
After that, we parsed its privacy policy and identified whether it explicitly mentioned log-related terms (e.g., “log” and “logging” ). To ensure detection accuracy, we conducted a manual inspection of the extracted matches. This step was necessary because certain substrings, such as ``catalog,'' ``login,'' or ``logic'' contain the term ``log'' but are unrelated to logging.
The resulting binary indicator (presence or absence of log-related disclosure) served as the dependent variable for our analysis.

To investigate which factors influence the presence of privacy policies in Android applications, we conducted a statistical analysis based on app-level metadata collected from the Google Play Store. Each application in our dataset was annotated with attributes including categories, downloads, user ratings and the number of reviews.

First, we examined whether the likelihood of providing a privacy policy varied across different app categories. To assess this association, we performed a \textit{Chi-square test}~\cite{chisquare} of independence, which allowed us to evaluate whether the distribution of privacy policy presence was significantly related to app category or downloads.


Second, to explore whether user engagement indicators were correlated with the presence of privacy policies, we compared continuous variables such as average user ratings and number of reviews between apps with and without log-related privacy policies. Since the variances between groups were unequal and sample sizes differed, we employed \textit{Welch’s t-statistic}~\cite{west2021best} to test for significant differences in means.

\textbf{\textit{Results.}}
In the following, we present how different app metadata influence the disclosure of logging practices in privacy policies.

\subsubsection{Application categories}
We first examined whether the disclosure of logging practices in privacy policies varied across application categories.
The Chi-square test resulted in a statistic of $\chi^2(41) = 52.53$, $p = 0.107$, indicating no statistically significant association between app category and the likelihood of mentioning log-related policies.

\subsubsection{Download}
In contrast, when analyzing download volume, a significant relationship was observed. 
The Chi-square test produced $\chi^2(17) = 50.93$, $p < 0.001$, showing that applications with higher download volumes were significantly more likely to include log-related statements in their privacy policies.
Specifically, among the 154 apps with fewer than 100 thousand downloads, only 27 (17.5\%) mentioned logging; this proportion increased to 100 (25.6\%) among the 391 apps with 100 thousand to 5 million downloads; and reached 158 (34.73\%) among the 455 apps with more than 5 million downloads.



\subsubsection{Rate}

Prior to analysis, applications with missing rating values were excluded to ensure valid group comparisons.
The Welch’s \textit{t}-test revealed no statistically significant difference in average user ratings between applications with and without a log-related valid privacy policy,
$t(503.3) = -1.24, p = 0.217$. 
Applications without a valid privacy policy 
($M = 4.31$, $SD = 0.61$)
and those with a valid log-related privacy policy 
($M = 4.26$, $SD = 0.56$)
showed comparable rating distributions. 
The negative \textit{t}-value indicates that, on average, applications with a valid log-related privacy policy had slightly lower ratings than those without, but this difference was not statistically meaningful.

\subsubsection{Reviews}
Since some applications did not report review counts, those with missing review data were excluded from the analysis.
A \textit{Welch’s t-test} was conducted to compare the logarithm of user review counts between applications with and without a valid log-related privacy policy that explicitly mentioned log practices. 
Due to the extremely large and discontinuous distribution of review counts, the log transformation ($\log_{10}$) was applied to the review counts to correct for skewness and reduce the influence of extreme values, improving the plausibility of normality and accommodating unequal variances across groups.
Applications incorporating a log-related valid privacy policy ($M = 5.23$, $SD = 1.45$) exhibited substantially higher log-transformed review counts than those lacking such a policy ($M = 4.78$, $SD = 1.59$).
The results of the Welch’s \textit{t}-test are
$t(820.3) = 5.26, p < 0.001$.
This indicates that the observed difference in log-transformed review counts between the two groups is highly significant, with a negligible probability of occurring by chance. The degrees of freedom (820.3) reflect Welch’s adjustment for unequal variances, and the large positive \textit{t}-value suggests that apps with valid privacy policies receive substantially more user reviews.

\begin{Summary}{Summary of RQ1}{}
The results show that app metadata have a partial influence on the disclosure of logging practices. App category and user rating show no significant relationship with disclosure behavior. In contrast, apps with higher download volumes and more user reviews are significantly more likely to include log-related statements in their privacy policies. These findings suggest that app popularity is positively associated with transparency in disclosing logging activities.
\end{Summary}

\subsection*{\textbf{RQ2: How do privacy policies specify the logging practices of Android applications?}}





\textbf{\textit{Motivation.}}
Prior research~\cite{ZimmeckWZILSWSB16,WilsonSDLCLAZSR16,YuLLZ16} have shown that many privacy policies provide ambiguous or incomplete information about data collection and processing. But rarely has research shown the connection between privacy policies and logging practice. It remains unclear how explicitly Android apps describe their logging practices, i.e., what is logged, why it is logged, and how much detail is provided to users.

By systematically analyzing how privacy policies specify logging practices, this study aims to assess the degree of transparency and granularity in such disclosures. The findings can help determine whether current policies meaningfully inform users about logging behaviors or rely on generalized statements that obscure privacy-relevant details, thereby providing insights into improving transparency and accountability within the Android ecosystem.

\textbf{\textit{Approach.}} To investigate how Android applications specify their logging practices in privacy policies, we designed a mixed-method workflow combining manual annotation, reference-based GPT-assisted classification, and human validation. The analysis focuses on three key dimensions: 
(1) what information is logged, (2) why logs are collected, and (3) how logging practices are described.

We first manually analyzed a subset of privacy policy statements to establish representative categories, detection keywords, contextual indicators, and example phrases for each analysis dimension. This manual phase resulted in a high-quality reference set that captures the linguistic and contextual patterns of log-related disclosures. The curated references were then used as prompt-based exemplars to guide the LLM in automatically classifying the remaining dataset. In this study, we employed GPT-5 to help analyze logging practice statements in privacy policies, as such disclosures are often expressed using heterogeneous and context-dependent language that is difficult to capture using fixed keyword rules alone.

After automatic classification, two independent reviewers examined and corrected the LLM-generated labels to ensure accuracy and consistency.  Disagreements were resolved through discussion, and the final merged results formed the validated dataset used for subsequent quantitative analysis.

\textbf{\textit{Results.}}
We found and analyzed 285 logging practice statements in privacy policies, supplemented by human verification. 
The following presents the results of RQ2:

\setcounter{subsubsection}{0}
\subsubsection{What is logged — Content Categorization}

\begin{table}[ht]
\footnotesize
\centering
\caption{Categories of Logged Information} \label{tab:whatlog}    
\begin{tabular}{@{}p{6em}p{13em}p{8em}p{3em}@{}}
\toprule
\textbf{Category} & \textbf{Definition} &  \textbf{Examples} &  \textbf{App}\\
\midrule
System Information & Information of the user's device or software. & operating system version, browser type, hardware details, and network information & 214 (30.27\%) \\ \midrule
Usage logs & Data describing how users interact with the application. & page visits, clicks, search activities, and other behavioral traces & 195 (27.58\%) \\ \midrule
Identifiers  & Information used to uniquely identify users or devices. This information is often related to account linking, tracking, or personalization. &  IP addresses, device IDs, cookies, or advertising identifiers & 160 (22.63\%) \\ \midrule
Diagnostic Data  & Logs generated for stability and maintenance purposes. & crash reports, performance metrics, latency measurements, or error traces & 70 (9.9\%) \\ \midrule
Communication logs & Information related to user communications or interactions through the app. & call history, message data, phone numbers, or contact information & 29 (4.1\%) \\ \midrule
Uncategorized & Vague or generic statements lack sufficient context to determine the specific type of logged content. & technical data, information & 26 (3.68\%) \\ \midrule
Location Information & Geographic data collected from GPS, Wi-Fi, or network-based positioning to provide location-based services or analytics.  & gps, location & 13 (1.84\%)  \\ 
\bottomrule
\end{tabular}
\end{table}

For each log-related statement, we identified what types of information were claimed to be logged.
According to Table~\ref{tab:whatlog}, system information (30.27\%) is the most frequently declared type of data mentioned in Android app privacy policies, suggesting that most applications tend to log technical details. Usage logs are also highly prevalent (27.58\%), indicating that many apps monitor user interactions and feature usage. In addition, identifiers are commonly collected (22.63\%), reflecting widespread practices of user or device tracking. In contrast, diagnostic data (9.9\%) and communication logs (4.1\%) are mentioned far less often.
The relatively small number of uncategorized cases indicates that most privacy policies explicitly specify the types of information being logged, including system details, user activity data, or identifiers. Nevertheless, the existence of a few non-specific statements reveals inconsistencies in disclosure practices, which may hinder users’ ability to fully understand the extent of log data collection.

Additionally, the results show that most privacy policies mention multiple categories of logged information, rather than focusing on a single category. Specifically, 98 policies describe three categories of logged data, while 81 refer to only one category, and smaller portions mention two, four, or five categories. This suggests that many applications engage in diverse logging practices. However, the fact that a considerable number of policies refer to only one or two categories indicates that some developers provide limited disclosure of their logging scope.

\subsubsection{Why Logs Are Collected — Purpose Categorization}
We then analyzed the stated purpose of collecting logs. 
Table~\ref{tab:whylog} reveals that analytics (27.73\%) is the primary purposes for log collection in Android app privacy policies, suggesting that most applications use logs to monitor usage patterns, analyze user behavior, and optimize performance. 
Debugging and maintenance (24.27\%) are also frequently mentioned, while the \textit{What is logged} results show that diagnostic data are mentioned far less often.
This mismatch indicates that developers may collect or rely on diagnostic logs but choose not to explicitly describe them in privacy policies, revealing a possible lack of transparency in technical data disclosure. 
Additionally, a considerable proportion of policies remain unspecified (24\%) regarding the purpose of log collection, highlighting a lack of transparency in data disclosure. Targeting (9.33\%) purpose appears less frequently but are still present, indicating that some applications utilize logs for commercial analytics and targeted recommendations. Meanwhile, security and protection are mentioned relatively infrequently (8.53\%), implying that only a small number of policies explicitly acknowledge the use of logs for security or abuse detection purposes.

\begin{table}[ht]
\footnotesize
\centering
\caption{Categories of Logging Purposes} \label{tab:whylog}    
\begin{tabular}{@{}p{6.5em}p{13em}p{8em}p{3em}@{}}
\toprule
\textbf{Category} & \textbf{Definition} &  \textbf{Examples}&  \textbf{App}\\ \midrule
Analytics & Logs collected to analyze user behavior, usage trends, or overall app performance. Such logs are typically aggregated and used to measure engagement or improve user experience. & We analyze usage data to understand feature popularity. & 104 (27.73\%) \\ \midrule
Debugging and Maintenance & Logs collected to support debugging, error diagnosis, or app maintenance. & crash logs, help request &  91 (24.27\%) \\ \midrule
Unspecified & Vague references to log collection without a clearly stated purpose. & We may collect log data as necessary. & 90 (24\%) \\ \midrule
Targeting & Logs used for advertising, personalization, or recommendation purposes. They are commonly linked to monetization or content customization. & We use your activity data to deliver personalized ads. & 35  (9.33\%)  \\ \midrule
Protection & Logs collected for security, fraud prevention, or abuse detection. These statements emphasize safeguarding users or systems from malicious or unauthorized activities & We collect IP addresses to prevent fraud. & 32  (8.53\%) \\ \midrule
Compliance & Logs collected to satisfy legal obligations, regulatory requirements, or official investigations. & Logs may be retained for legal reasons. &  13 (3.47\%) \\ \midrule
Service Provision & Data that is necessary to deliver, operate, or support the core functions of a service requested by the user. & We store your uploaded files so you can access them later. & 10 (2.67\%) \\ 
\bottomrule
\end{tabular}
\end{table}

\subsubsection{How Logs Are Described — Specificity Level}

We analyzed how clearly the logging practices were described. Table~\ref{tab:howlog} reveals that the majority of privacy policies (72.28\%) provide a high level of specificity when discussing log collection, with most statements offering relatively detailed explanations of what is logged and for what purpose. This indicates that developers generally tend to include clear and explicit descriptions of their logging practices. However, a smaller number of policies were classified as having medium (17.19\%) or low (10.53\%) specificity, suggesting that some developers still rely on vague or generalized language such as ``we may collect log information'' without clarifying the exact scope or nature of the data involved. These less specific disclosures may limit users’ ability to fully understand how their information is handled.

\begin{table}[ht]
\footnotesize
\centering
\caption{Categories of Specificity Level} \label{tab:howlog}    
\begin{tabular}{@{}p{5em}p{13em}p{11em}@{}}
\toprule
\textbf{Category} & \textbf{Definition} &  \textbf{Examples}\\ \midrule
High (72.28\%) & Concrete description of data types or processes. & Crash logs include OS version and app ID.\\ \midrule
Medium (17.19\%) & Abstract but still interpretable. & We collect usage data to improve performance.\\ \midrule
Low (10.53\%) & Vague references with minimal detail & certain data, some information \\
\bottomrule
\end{tabular}
\end{table}




\begin{Summary}{Summary of RQ2}{}
The results show that most Android app privacy policies mentioning logging practices clearly describe what is logged and why, mainly focusing on system information and analytics. However, some policies remain vague or incomplete, indicating inconsistent transparency in how logging practices are disclosed.
\end{Summary}

\subsection*{\textbf{RQ3: To what extent do log-related statements in privacy policies align with the privacy disclosures present in real-world Android application logs?}}

\textbf{\textit{Motivation.}}
While privacy policies present how applications claim to handle user data, they often represent only the declared dimension of privacy practices. In reality, Android applications continuously generate logs that capture sensitive information. Discrepancies between what is disclosed in policies and what is actually logged can expose significant transparency and compliance issues.
If applications record sensitive data without corresponding disclosures, users are left unaware of potential privacy risks, and regulators cannot reliably assess adherence to data protection requirements. Evaluating the alignment between policy statements and real-world logging behaviors is therefore essential for assessing the credibility of privacy disclosures and the integrity of Android’s data ecosystem.
To this end, this research examines to what extent the log-related statements in privacy policies align with the privacy-relevant content observed in real-world application logs.







\textbf{\textit{Approach.}}
Our approach consistes of two stages:

\setcounter{subsubsection}{0}
\subsubsection{Sensitive Information Detection}
We applied a four-phase strategy for detecting sensitive information in Android app logs.
Traditional rule-based approaches, which rely on predefined keywords or regular expressions, are often insufficient for this task, as sensitive information may be logged using diverse and context-dependent expressions which causes cases missing~\cite{ZhiyuanIEEE}. 
We employed Brain as the log parsing component in our pipeline to handle the scale and redundancy of Android application logs. The collected log files contain a large number of repeated log lines, which makes analysis inefficient and unnecessary.


\begin{itemize}
    \item \textit{Seed Keyword List.} We began with a seed list of sensitive information keywords, which was derived from prior work~\cite{ZhiyuanPoster}. This list included keywords that are commonly used to capture privacy-relevant data, such as personal identifiers (e.g., name, IP, email).
    \item \textit{Keyword Expansion via Pilot Study.} 
    To extend the initial list of sensitive information keywords, we conducted a pilot study on logs collected from a subset of Android applications.
    We opted not to inspect every log entry across all apps for two reasons:
    (1) \textit{uniformity}, as our dataset was self-constructed and all applications were executed under the same experimental conditions, so the types of potentially exposed personal information were identical across apps;
    and (2) \textit{the size of the final dataset}, since the final dataset was sufficiently large that exhaustive manual inspection would have been infeasible within a reasonable timeframe.
    We therefore selected a subset of applications using a random sampling strategy. Specifically, each application in the dataset was assigned a unique identifier, and a pseudo-random number generator was used to select applications without replacement. The sampling size was determined based on a 95\% confidence level, resulting in 278 selected applications out of the 1,000 available apps. 
    
    After that, we further filtered the sampled logs to obtain a manageable subset that could be manually examined within a feasible timeframe.
    Each selected log file was subsequently processed using \textit{Brain}~\cite{Siyu2023}, a log parsing tool designed to automatically extract log templates and cluster similar log lines into event groups.
    We chose \textit{Brain} because of its superior capability in processing Android application logs~\cite{zhu2023loghub, jiang2024loghub2}.
    Since log entries within the same event group share the same log template and typically exhibit similar privacy leakage patterns, we retained only one representative log entry per group.
    This strategy significantly reduced the verification workload while maintaining confidence in the accuracy of our detection.

    As a result, the 278 representative log files were merged into a single dataset and analyzed using ChatGPT (GPT-5) through few-shot prompting. The model was instructed to identify potential sensitive information patterns and propose additional candidate keywords not present in the initial seed list. These suggested keywords were then manually reviewed and validated to ensure relevance and accuracy before being incorporated into the final keyword set. The prompt for detecting potential sensitive information could be found in the repository.
    

    \item \textit{Detection.} We implemented a Python-based detection script that applied the enhanced keyword list to the collected log files. 
    Each log line was then compared against both the keyword list and a set of regular expression patterns, derived from the same list, designed to detect structured sensitive data such as IP addresses and GPS coordinates.
    \item \textit{Manual Inspection.} To validate the accuracy of our detection results, we conducted a manual inspection. This step was necessary to confirm that the matches truly corresponded to sensitive information rather than false positives. For example, place names may indicate sensitive location information, but in some cases, they refer to the name of an organization.

    
\end{itemize}

\subsubsection{Alignment Between Privacy Policies and Privacy Disclosures in Logs}
We collected the privacy policies associated with each application from the Google Play Store or official developer websites.
To assess whether applications’ privacy policies align with their real-world logging practices, we use two metrics:



\begin{itemize}
    \item \textit{Omission:}
Applications leak sensitive information in logs but either have no privacy policy, have a policy that makes no reference to logging practices, or disclose logging practices in their privacy policies without mentioning the specific types of leaked information.


    \item \textit{Inconsistency:}
Although applications explicitly mention their logging practices in privacy policies, the actual log contents still reveal privacy leakages that are inconsistent with the purposes stated in those policies.
\end{itemize}


\textbf{\textit{Results.}}
In the following, we present the results for RQ3, which are organized into two parts:

\setcounter{subsubsection}{0}
\subsubsection{Sensitive Information Detection}

Our initial seed keyword list primarily covered device-related identifiers (e.g., manufacturer, model, Android ID, serial number, IMEI), user identifiers (e.g., email, name), location information (e.g., GPS coordinates, fine-grained city, coarse-grained region), and network identifiers (e.g., IP address). 
During detection, we observed several limitations of an initial keyword-based list.
\begin{itemize}
    \item \textit{Incomplete coverage:} Certain user identifiers, such as names and email addresses, were not included in the initial privacy information list but were still detected in the logs. Interestingly, some of these identifiers did not belong to the test accounts used in our experiments. Instead, they appeared to be residual system data left over from previous activities or cached records. In addition, we identified instances of Wi-Fi network names being logged, which were not included in the seed keyword list and revealed the name of our institution.
    \item \textit{Format variability:} Sensitive information such as GPS coordinates and IP addresses appeared in multiple formats. For GPS values, the integer part before the decimal point is relatively stable, while the fractional part after the decimal point may vary in length. This variability makes exact keyword matching ineffective. 
    To address these problems, we employed regular expressions to capture numerical patterns of GPS coordinates.
\end{itemize}


We added six additional email addresses and one additional name to the initial privacy keyword list, resulting in a total of seven email addresses and two names. And we improved detection of GPS coordinates and IP addresses using regular expressions. 
Table~\ref{tab:table_privacy_info} lists the final categories of privacy information identified in our analysis.

\begin{table}[ht]
\small
\centering
\caption{A list of the privacy keywords for privacy leakage detection.} \label{tab:table_privacy_info}    
\begin{tabular}{@{}lp{5cm}@{}}
\toprule
\textbf{Info. Type} & \textbf{Value}\\
\midrule
Manufacturer / model    &  Google Pixel 5a  \\
Android ID     &  1169...8f73e  \\
AndroidID (SHA-1) & 4b7a...b752 \\
Android ID (MDS)    &  8f0d...e4dd5 	\\         
Email    &  rittest10@gmail.com, ritcs2022@gmail.com and other five private emails\\ 
Serial    &  1703...0062	\\         
Serial (MDS)    &  c659...0c75 	\\         
Name    &  Kirito Kirigaya, RIT 	\\         
IMEI    &  3576...99347 / 63	\\         
GPS    &  43.xxx, -77.xxx 	\\         
Fine-grained location    &  Rochester 	\\         
Coarse-grained location    &  New York 	\\         
IP    &  Two IP addresses	\\            
\bottomrule
\end{tabular}
\end{table}



\begin{figure}[h]
    \centering
    \includegraphics[width=0.95\columnwidth]{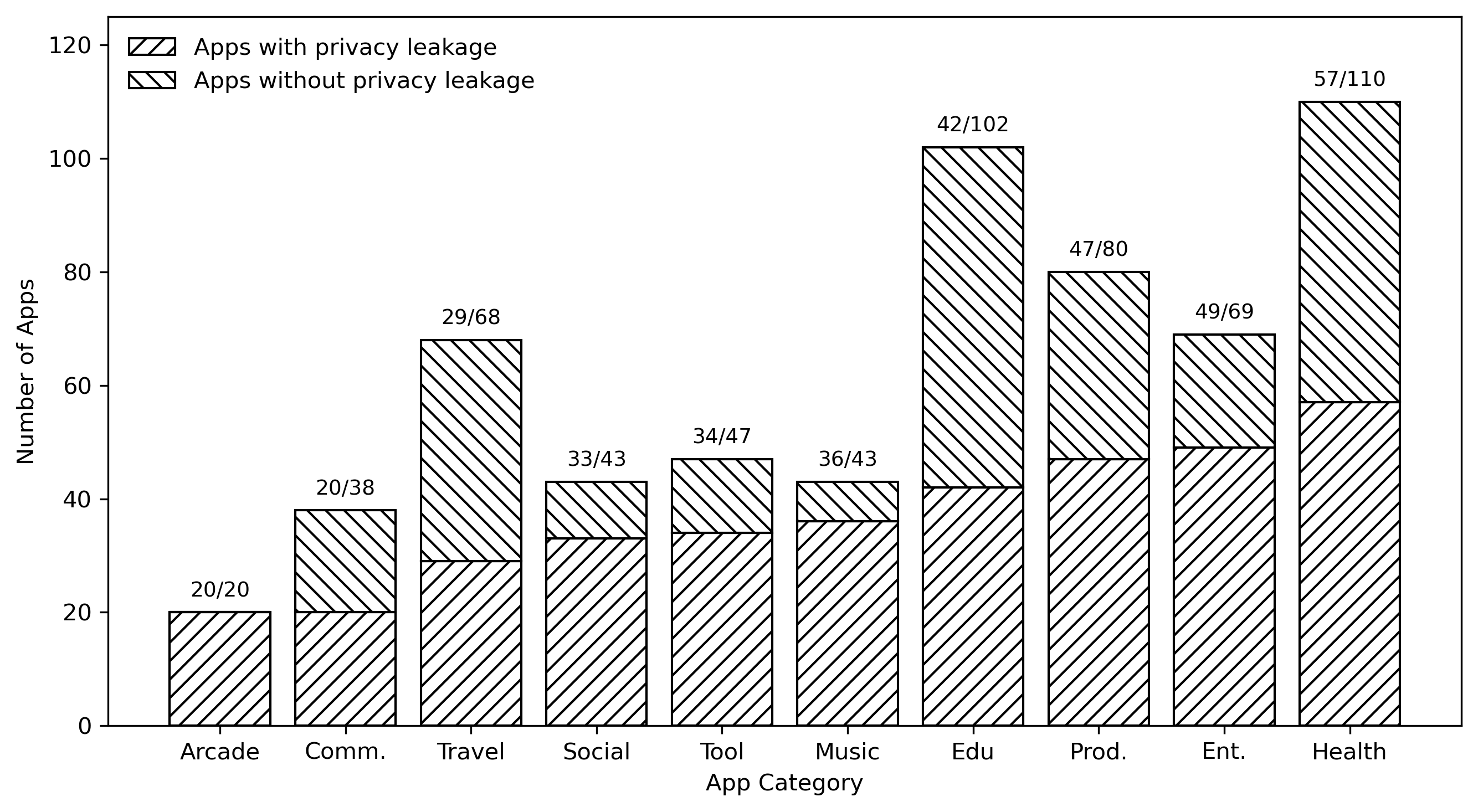}
    \caption{Top-10 App Categories with Detected Privacy Leakage, numbers above bars are shown as apps with privacy leakage / total apps.}
    \label{fig:table_app_category}
\end{figure}

In total, we analyzed 1,000 Android applications. Among them, 607 applications (60.7\%) were found to contain privacy leakages in their logs. Figure~\ref{fig:table_app_category} presents the distribution of these 1,000 applications across categories. 
We observed that Health \& Fitness (57 apps), Entertainment (49 apps), Productivity (47 apps), and Education (42 apps) are among the most represented categories. These domains often handle sensitive or user-specific data, which may explain their higher representation in terms of privacy policy disclosures.
Conversely, other categories (240 apps) like comics, food services, and racing contain only a handful of apps in our dataset have privacy leakage issue, these domains may place relatively less emphasis on developing or disclosing privacy policies.


\subsubsection{ Alignment Between Privacy Policies and Privacy Disclosures in Logs}

    In total, 715 out of 1,000 applications lacked any log-related disclosure in their privacy policies.
    These findings indicate that log-related practices are commonly overlooked in existing privacy policies, reflecting significant gaps in transparency and policy integrity.

Table \ref{tab:table_privacy_leakage_policy_1000} summarizes the alignment between the privacy policies and the sensitive information identified in application logs.
A total of 393 applications showed no privacy leakage detected in their logs.
In 4 cases, the privacy information leaked in the logs was included in the policy statements, meaning that most privacy policies lack completeness and specificity in their disclosures.
Among the 1,000 analyzed applications, 603 exhibited privacy leakage cases where the leaked information type was not mentioned in the policy.
For example, the app states in its policy: ``We may engage a data provider who may collect web log data from you (including IP address  and information about your browser or operating system), or place or recognize a unique cookie on your browser to enable you to receive customized ads or content.'' However, our analysis revealed privacy leaks involving information such as  manufacturer/model, email address, fine-grained location, and IP address, while only the IP address was explicitly mentioned in the policy.

These results reveal a substantial discrepancy between stated privacy policies and actual data handling behaviors, with the majority of applications either omitting or only partially addressing their log-related data collection practices.

\begin{table}[ht]
\small
\centering
\caption{Alignment between privacy leakages and privacy policy statements across 1,000 Applications.} \label{tab:table_privacy_leakage_policy_1000}    
\begin{tabular}{@{}lr@{}}
\toprule
\textbf{Type} & \textbf{Count}\\
\midrule
Not mentioned in the policy    &  603  \\
No privacy leakage    &  393  \\
Include & 4 \\
\midrule
Total   &   1,000 \\
\bottomrule
\end{tabular}
\end{table}



Table~\ref{tab:table_leakage_case_not_in_policy} summarizes the distribution of detected privacy leakage types across 1,000 analyzed applications. The first column lists the information categories found in execution logs, while the second column (\textit{Total Leakage}) indicates the total number of leakage instances identified. The third column (\textit{Unstated Leakage}) represents cases where the leaked information type was not disclosed in the corresponding privacy policy. As shown in the table, fine-grained location information was the most frequently leaked category, accounting for over half of all identified leakages (37,376 instances). IP addresses (11,311 instances) and device manufacturer/model identifiers (10,484 instances) were also prevalent. Among these, a large proportion remained undisclosed in privacy policies—for example, 76.34\% of IP leakages and 97.51\% of manufacturer/model leakages lacked corresponding log-related policy statements. These findings demonstrate that even when sensitive data types like location or device identifiers are routinely collected, most applications fail to acknowledge such practices in their privacy policies.

\begin{table}[ht]
\small
\centering
\caption{Total privacy leakage cases found in 1,000 apps and corresponding undisclosed cases where privacy policies failed to mention the leaked information types.} \label{tab:table_leakage_case_not_in_policy}    
\begin{tabular}{@{}lrrr@{}}
\toprule
\textbf{Info. Type} & \textbf{Total Leakage}& \textbf{Unstated leakage} & \textbf{App}\\
\midrule
IP    &  11,311  &   8,635   & 347\\
Manufacturer / model    &  10,484  & 10,223 & 551\\
Email    &  2,204  & 1,857 & 202\\
Fine-grained location & 37,376 &  26,479 & 234\\
Name    &  635  &   494 & 39\\
Coarse location    &  166  &    146 & 62\\
GPS    &  95  & 67 & 69\\
\midrule
Total   &   62,271 & 47,901 & 607\\
\bottomrule
\end{tabular}
\end{table}




\begin{Summary}{Summary of RQ3}{}
The results show that privacy leakages are common in Android applications, yet the leaked log-related data types are rarely mentioned in their privacy policies. Only a few apps demonstrated consistency between their policies and actual behavior, suggesting that transparency in log-related data handling is still largely insufficient.
\end{Summary}

\section{Discussion}\label{sec:discussion}
This section discusses the broader implications of our findings and highlights key challenges in aligning privacy policies with actual logging practices, and the recommendations for developers .

\subsection{The Accountability Gap in Popular Applications}
Our findings reveal an interesting paradox: while more popular applications (with higher download volumes and more user reviews) are significantly more likely to include log-related statements in their privacy policies, the overall disclosure rate remains strikingly low (28.5\%) across the Android ecosystem. This suggests that market success creates incentives for improved transparency, but these incentives remain insufficient to establish comprehensive disclosure as a standard practice.

Several factors might explain this pattern. First, popular applications face greater regulatory scrutiny and legal liability due to their larger user base, creating stronger incentives for disclosure. Second, applications with larger user bases likely receive more privacy-related inquiries and feedback, potentially driving more comprehensive policy development. Third, established applications typically have greater resources to invest in privacy compliance and legal expertise.

However, the continued low rate of disclosure among even highly popular applications suggests that market forces alone may be insufficient to ensure transparency. This accountability gap raises important questions about whether additional regulatory frameworks or platform-level requirements are needed to elevate privacy disclosure standards across the ecosystem, regardless of an application's popularity or market position.

\subsection{The Technical-Legal Disconnect}
A striking inconsistency in our findings is the mismatch between technical logging implementations and legal disclosures. While debugging and maintenance are cited as the important purposes for logging (24.27\%), diagnostic data is mentioned far less frequently (9.9\%) in descriptions of what is logged. This discrepancy suggests a fundamental disconnect between the technical teams implementing logging functionality and the legal teams drafting privacy policies.

This disconnect likely stems from organizational silos where development teams may not effectively communicate with privacy policy authors about what data is actually being collected. Software engineers often implement logging primarily as a technical necessity, without considering its privacy implications or disclosure requirements. Meanwhile, legal teams may draft policies based on generic templates or regulatory requirements, without detailed knowledge of specific implementation details.

The challenge is further complicated by the dynamic nature of mobile application development. Logging practices often evolve through iterative development cycles, with new log statements being added as features are implemented or bugs are identified. Unless there are robust processes for continuously updating privacy policies to reflect these changes, disclosures will inevitably lag behind actual practices.

To address this disconnect, organizations need more integrated approaches to privacy management that bridge technical implementation and legal disclosure. This might include privacy-aware logging frameworks that automatically document what information is being captured, regular privacy impact assessments that review logging practices, and better communication channels between development and legal teams.

\subsection{The Detection Challenge in Modern Applications}

Our methodology revealed key limitations of traditional keyword-based approaches to detecting privacy leakages. LLM-assisted analysis expanded the initial keyword list and uncovered how modern applications often employ non-obvious data patterns and variable formats that evade simple detection.

Sensitive information such as GPS coordinates and IP addresses appeared in diverse formats, and residual system data unrelated to test accounts were also detected—indicating potential privacy leakages beyond what conventional methods capture. This suggests that the true extent of privacy risks in mobile applications may be underestimated.

These findings highlight the need for more adaptive detection techniques capable of recognizing sensitive information across varied contexts. Future approaches may combine keyword-based and machine learning methods or adopt privacy-aware logging frameworks to automatically flag potential leakages during development.


\subsection{Recommendations}

\textbf{Recommendations for Developers.}
Our findings suggest that developers should treat logging data as a first-class data source subject to the same transparency and minimization principles as primary data flows. Privacy policies should explicitly disclose what types of data are logged, for what purposes (e.g., debugging, analytics), and whether such data are shared with third parties. Moreover, developers should avoid logging sensitive identifiers (e.g., user IDs, device IDs, or exact location) unless strictly necessary and apply retention limits to logging data. Technical measures, such as local filtering, anonymization, and end-to-end encryption for transmitted logs, should be implemented to minimize privacy risks. Regular audits of logging practices should be integrated into the software development lifecycle to ensure ongoing compliance with stated policies.

\textbf{Recommendations for Policymakers and Regulators.}
Our study highlights logging as an under-examined channel of personal data collection that is often insufficiently disclosed in privacy policies. 
Regulators may consider issuing clearer guidance on the treatment of logging and diagnostic data under existing privacy frameworks (such as GDPR or CCPA), clarifying when such data constitutes personal data and what obligations apply.

\textbf{Implications for App Users.}
Our results indicate that the absence of logging-related disclosures limits the effectiveness of privacy policies as a decision-making tool for users. Users should be aware that even when privacy policies appear comprehensive, logging and diagnostic data collection may still occur without explicit disclosure. This underscores the need for greater caution when granting permissions and relying solely on privacy policies to assess an app’s data practices.

\section{Threats to Validity} \label{sec:threat}


\noindent \textbf{Internal Validity.}
Internal validity concerns whether the findings accurately reflect the analyzed phenomena. To minimize bias, all apps were tested under consistent experimental conditions, including identical device settings and interaction durations. Manual inspection was applied in multiple phases, such as the validation of privacy policy parsing and sensitive information detection, to verify automated results. However, the manual steps may still introduce subjective bias in classification and labeling. Moreover, while the LLM-assisted keyword expansion improved coverage, false positives or negatives may still occur in sensitive data detection due to context ambiguity.

\noindent \textbf{External Validity. }
The study selected 1,000 Android applications from diverse categories and popularity levels to improve representativeness. This selection strategy helps capture different types of apps and developer practices. However, the dataset still cannot represent the entire Android ecosystem. Paid applications and region-specific apps that are not distributed through Google Play remain outside our analysis. The testing process also relied on manual interaction, and each app was operated following a predefined set of actions. These interactions may not cover all user behaviors or runtime conditions that influence how an app generates logs. 

\noindent \textbf{Construct Validity.}
We defined log-related disclosure and privacy leakage based on established research, but these definitions still involve interpretation. Some types of information recorded in logs, such as device models, may not always represent privacy leakage depending on how the data are used. Likewise, when a privacy policy mentions log collection, it does not necessarily mean that the statement provides clear or complete information. In several cases, policies stated that location data would be recorded in logs but did not specify whether the location referred to precise GPS data or coarse network-based location. This lack of clarity makes it difficult to determine how well the policy descriptions align with the actual logging practices. Such ambiguity in meaning and level of detail can affect the accuracy of our alignment analysis and, consequently, the validity of our results.

\section{Conclusion and Future Work} \label{sec:conc}


This study revealed patterns of log disclosure in Android privacy policies, as well as significant gaps between policy statements and actual logging behaviors in Android applications. Our analysis of 1,000 Android apps found that most provide privacy policies, yet very few explicitly mention logging practices despite their prevalent use. Popular applications were more likely to disclose logging practices.
Among the policies that reference logging, most clearly describe what information is logged; however, 27.7\% of log-related statements remain overly simplistic or vague, offering limited insight into actual data collection activities. We also found widespread privacy leakages in application logs, with most apps leaking information not mentioned in their policies.
The alarming misalignment between declared and actual practices reveals a transparency crisis in the Android ecosystem that undermines user trust and potentially violates regulatory requirements. These findings highlight that current privacy policies provide incomplete or ambiguous descriptions of logging practices, which frequently do not align with actual logging behaviors. Our future work will expand the dataset to include paid and region-specific applications, explore automated methods for contextual log analysis, and investigate how developers interpret regulatory requirements when drafting privacy policies.



\bibliographystyle{ACM-Reference-Format}
\bibliography{reference}

@INPROCEEDINGS{ZhiyuanPoster,
  author={Chen, Zhiyuan and Deo, Soham Sanjay and Puttaparthi, Poorna Chander Reddy and Tang, Yiming and Zhang, Xueling and Shang, Weiyi},
  booktitle={2024 39th IEEE/ACM International Conference on Automated Software Engineering (ASE)}, 
  title={From Logging to Leakage: A Study of Privacy Leakage in Android App Logs}, 
  year={2024},
  volume={},
  number={},
  pages={2484-2485},
  }

@inproceedings{ZimmeckWZILSWSB16,
  author       = {Sebastian Zimmeck and
                  Ziqi Wang and
                  Lieyong Zou and
                  Roger Iyengar and
                  Bin Liu and
                  Florian Schaub and
                  Shomir Wilson and
                  Norman M. Sadeh and
                  Steven M. Bellovin and
                  Joel R. Reidenberg},
  title        = {Automated Analysis of Privacy Requirements for Mobile Apps},
  booktitle    = {2016 {AAAI} Fall Symposia, Arlington, Virginia, USA, November 17-19,
                  2016},
  publisher    = {{AAAI} Press},
  year         = {2016},
}

@article{ZimmeckSSRWRRS19,
  author       = {Sebastian Zimmeck and
                  Peter Story and
                  Daniel Smullen and
                  Abhilasha Ravichander and
                  Ziqi Wang and
                  Joel R. Reidenberg and
                  N. Cameron Russell and
                  Norman M. Sadeh},
  title        = {{MAPS:} Scaling Privacy Compliance Analysis to a Million Apps},
  journal      = {Proc. Priv. Enhancing Technol.},
  volume       = {2019},
  number       = {3},
  pages        = {66--86},
  year         = {2019},
  url          = {https://doi.org/10.2478/popets-2019-0037},
  doi          = {10.2478/POPETS-2019-0037},
}

@inproceedings{WilsonSDLCLAZSR16,
  author       = {Shomir Wilson and
                  Florian Schaub and
                  Aswarth Abhilash Dara and
                  Frederick Liu and
                  Sushain Cherivirala and
                  Pedro Giovanni Leon and
                  Mads Schaarup Andersen and
                  Sebastian Zimmeck and
                  Kanthashree Mysore Sathyendra and
                  N. Cameron Russell and
                  Thomas B. Norton and
                  Eduard H. Hovy and
                  Joel R. Reidenberg and
                  Norman M. Sadeh},
  title        = {The Creation and Analysis of a Website Privacy Policy Corpus},
  booktitle    = {Proceedings of the 54th Annual Meeting of the Association for Computational
                  Linguistics, {ACL} 2016, August 7-12, 2016, Berlin, Germany, Volume
                  1: Long Papers},
  publisher    = {The Association for Computer Linguistics},
  year         = {2016},
  url          = {https://doi.org/10.18653/v1/p16-1126},
  doi          = {10.18653/V1/P16-1126},
}

@article{liu2022evaluating,
  title={Evaluating the privacy policy of android apps: a privacy policy compliance study for popular apps in China and Europe},
  author={Liu, Kaijun and Xu, Guoai and Zhang, Xiaomei and Xu, Guosheng and Zhao, Zhangjie},
  journal={Scientific Programming},
  volume={2022},
  number={1},
  pages={2508690},
  year={2022},
  publisher={Wiley Online Library}
}

@online{GDPR,
	title        = {General Data Protection Regulation},
	author       = {GDPR},
	url          = {https://gdpr-info.eu/},
	urldate      = {2025-2-14},
}

@online{CCPA,
	title        = {California Consumer Privacy Act (CCPA)},
	author       = {State of California Department of Justice},
	url          = {https://oag.ca.gov/privacy/ccpa},
	urldate      = {2025-2-14},
}

@inproceedings{StoryZS18,
  author       = {Peter Story and
                  Sebastian Zimmeck and
                  Norman M. Sadeh},
  editor       = {Manel Medina and
                  Andreas Mitrakas and
                  Kai Rannenberg and
                  Erich Schweighofer and
                  Nikolaos Tsouroulas},
  title        = {Which Apps Have Privacy Policies? - An Analysis of Over One Million
                  Google Play Store Apps},
  booktitle    = {Privacy Technologies and Policy - 6th Annual Privacy Forum, {APF}
                  2018, Barcelona, Spain, June 13-14, 2018, Revised Selected Papers},
  series       = {Lecture Notes in Computer Science},
  volume       = {11079},
  pages        = {3--23},
  publisher    = {Springer},
  year         = {2018},
  url          = {https://doi.org/10.1007/978-3-030-02547-2\_1},
  doi          = {10.1007/978-3-030-02547-2\_1},
}

@inproceedings{YuLLZ16,
  author       = {Le Yu and
                  Xiapu Luo and
                  Xule Liu and
                  Tao Zhang},
  title        = {Can We Trust the Privacy Policies of Android Apps?},
  booktitle    = {46th Annual {IEEE/IFIP} International Conference on Dependable Systems
                  and Networks, {DSN} 2016, Toulouse, France, June 28 - July 1, 2016},
  pages        = {538--549},
  publisher    = {{IEEE} Computer Society},
  year         = {2016},
  url          = {https://doi.org/10.1109/DSN.2016.55},
  doi          = {10.1109/DSN.2016.55}
}

@ARTICLE{Siyu2023,
  author={Yu, Siyu and He, Pinjia and Chen, Ningjiang and Wu, Yifan},
  journal={IEEE Transactions on Services Computing}, 
  title={Brain: Log Parsing With Bidirectional Parallel Tree}, 
  year={2023},
  volume={16},
  number={5},
  pages={3224-3237},
  doi={10.1109/TSC.2023.3270566}}

@Inbook{chisquare,
author="Tallarida, Ronald J.
and Murray, Rodney B.",
title="Chi-Square Test",
bookTitle="Manual of Pharmacologic Calculations: With Computer Programs",
year="1987",
publisher="Springer New York",
address="New York, NY",
pages="140--142",
isbn="978-1-4612-4974-0",
doi="10.1007/978-1-4612-4974-0_43",
url="https://doi.org/10.1007/978-1-4612-4974-0\_43"
}

@article{Yousra2024,
author = {Javed, Yousra and Sajid, Ayesha},
title = {A Systematic Review of Privacy Policy Literature},
year = {2024},
issue_date = {February 2025},
publisher = {Association for Computing Machinery},
address = {New York, NY, USA},
volume = {57},
number = {2},
issn = {0360-0300},
url = {https://doi-org.ezproxy.rit.edu/10.1145/3698393},
doi = {10.1145/3698393},
journal = {ACM Comput. Surv.},
month = nov,
articleno = {45},
numpages = {43}
}

@INPROCEEDINGS{Chang2020,
  author={Chang, Kai Chih and Nokhbeh Zaeem, Razieh and Barber, K. Suzanne},
  booktitle={2020 Second IEEE International Conference on Trust, Privacy and Security in Intelligent Systems and Applications (TPS-ISA)}, 
  title={Is Your Phone You? How Privacy Policies of Mobile Apps Allow the Use of Your Personally Identifiable Information}, 
  year={2020},
  volume={},
  number={},
  pages={256-262},
  doi={10.1109/TPS-ISA50397.2020.00041}}

@misc{wagner2022privacypoliciesagescontent,
      title={Privacy Policies Across the Ages: Content and Readability of Privacy Policies 1996--2021}, 
        author={Isabel Wagner},
      year={2022},
      eprint={2201.08739},
      archivePrefix={arXiv},
      primaryClass={cs.CR},
      url={https://arxiv.org/abs/2201.08739}, 
}

@article{obar2020biggest,
  title={The biggest lie on the internet: Ignoring the privacy policies and terms of service policies of social networking services},
  author={Obar, Jonathan A and Oeldorf-Hirsch, Anne},
  journal={Information, Communication \& Society},
  volume={23},
  number={1},
  pages={128--147},
  year={2020},
  publisher={Taylor \& Francis}
}

@article{mcdonald2008cost,
  title={The cost of reading privacy policies},
  author={McDonald, Aleecia M and Cranor, Lorrie Faith},
  journal={Isjlp},
  volume={4},
  pages={543},
  year={2008},
  publisher={HeinOnline}
}

@article{baalous2025detecting,
  title={Detecting the Inconsistency between Android Apps’ Data Collection and Google Play’s Data Safety Using Static Analysis},
  author={Baalous, Rawan and Althobaiti, Alanoud and Alyoubi, Dareen and Alzahrani, Rama and Aljohani, Mona},
  journal={Cybernetics and Information Technologies},
  volume={25},
  number={1},
  year={2025}
}

@inproceedings {Benjamin2020,
author = {Benjamin Andow and Samin Yaseer Mahmud and Justin Whitaker and William Enck and Bradley Reaves and Kapil Singh and Serge Egelman},
title = {Actions Speak Louder than Words: {Entity-Sensitive} Privacy Policy and Data Flow Analysis with {PoliCheck}},
booktitle = {29th USENIX Security Symposium (USENIX Security 20)},
year = {2020},
isbn = {978-1-939133-17-5},
pages = {985--1002},
url = {https://www.usenix.org/conference/usenixsecurity20/presentation/andow},
publisher = {USENIX Association},
month = aug
}

@inproceedings{bui2021consistency,
  title={Consistency analysis of data-usage purposes in mobile apps},
  author={Bui, Duc and Yao, Yuan and Shin, Kang G and Choi, Jong-Min and Shin, Junbum},
  booktitle={Proceedings of the 2021 ACM SIGSAC Conference on Computer and Communications Security},
  pages={2824--2843},
  year={2021}
}

@inproceedings{khedkar2024advancing,
  title={Advancing android privacy assessments with automation},
  author={Khedkar, Mugdha and Schlichtig, Michael and Bodden, Eric},
  booktitle={Proceedings of the 39th IEEE/ACM International Conference on Automated Software Engineering Workshops},
  pages={218--222},
  year={2024}
}

@inproceedings{Yuxia2024,
author = {Zhan, Yuxia and Meng, Yan and Zhou, Lu and Xiong, Yichang and Zhang, Xiaokuan and Ma, Lichuan and Chen, Guoxing and Pei, Qingqi and Zhu, Haojin},
title = {VPVet: Vetting Privacy Policies of Virtual Reality Apps},
year = {2024},
isbn = {9798400706363},
publisher = {Association for Computing Machinery},
address = {New York, NY, USA},
url = {https://doi-org.ezproxy.rit.edu/10.1145/3658644.3690321},
doi = {10.1145/3658644.3690321},
booktitle = {Proceedings of the 2024 on ACM SIGSAC Conference on Computer and Communications Security},
pages = {1746–1760},
numpages = {15},
location = {Salt Lake City, UT, USA},
series = {CCS '24}
}

@inproceedings{Mukund2023,
author = {Srinath, Mukund and Sundareswara, Soundarya and Venkit, Pranav and Giles, C. Lee and Wilson, Shomir},
title = {Privacy Lost and Found: An Investigation at Scale of Web Privacy Policy Availability},
year = {2023},
isbn = {9798400700279},
publisher = {Association for Computing Machinery},
address = {New York, NY, USA},
url = {https://doi-org.ezproxy.rit.edu/10.1145/3573128.3604902},
doi = {10.1145/3573128.3604902},
booktitle = {Proceedings of the ACM Symposium on Document Engineering 2023},
articleno = {26},
numpages = {10},
location = {Limerick, Ireland},
series = {DocEng '23}
}

@article{Ding2024,
  author       = {Zishuo Ding and
                  Yiming Tang and
                  Xiaoyu Cheng and
                  Heng Li and
                  Weiyi Shang},
  title        = {\emph{LoGenText-Plus}: Improving Neural Machine Translation Based
                  Logging Texts Generation with Syntactic Templates},
  journal      = {{ACM} Trans. Softw. Eng. Methodol.},
  volume       = {33},
  number       = {2},
  pages        = {38:1--38:45},
  year         = {2024},
  url          = {https://doi.org/10.1145/3624740},
  doi          = {10.1145/3624740},
  bibsource    = {dblp computer science bibliography, https://dblp.org}
}

@inproceedings{Ding2023,
  author       = {Zishuo Ding and
                  Yiming Tang and
                  Yang Li and
                  Heng Li and
                  Weiyi Shang},
  title        = {On the Temporal Relations between Logging and Code},
  booktitle    = {45th {IEEE/ACM} International Conference on Software Engineering,
                  {ICSE} 2023, Melbourne, Australia, May 14-20, 2023},
  pages        = {843--854},
  publisher    = {{IEEE}},
  year         = {2023},
  url          = {https://doi.org/10.1109/ICSE48619.2023.00079},
  doi          = {10.1109/ICSE48619.2023.00079},
  bibsource    = {dblp computer science bibliography, https://dblp.org}
}

@article{Zhang2022,
  author       = {Haonan Zhang and
                  Yiming Tang and
                  Maxime Lamothe and
                  Heng Li and
                  Weiyi Shang},
  title        = {Studying logging practice in test code},
  journal      = {Empir. Softw. Eng.},
  volume       = {27},
  number       = {4},
  pages        = {83},
  year         = {2022},
  url          = {https://doi.org/10.1007/s10664-022-10139-0},
  doi          = {10.1007/S10664-022-10139-0},
  bibsource    = {dblp computer science bibliography, https://dblp.org}
}

@inproceedings{He2016,
  author       = {Pinjia He and
                  Jieming Zhu and
                  Shilin He and
                  Jian Li and
                  Michael R. Lyu},
  title        = {An Evaluation Study on Log Parsing and Its Use in Log Mining},
  booktitle    = {46th Annual {IEEE/IFIP} International Conference on Dependable Systems
                  and Networks, {DSN} 2016, Toulouse, France, June 28 - July 1, 2016},
  pages        = {654--661},
  publisher    = {{IEEE} Computer Society},
  year         = {2016},
  url          = {https://doi.org/10.1109/DSN.2016.66},
  doi          = {10.1109/DSN.2016.66},
  bibsource    = {dblp computer science bibliography, https://dblp.org}
}

@INPROCEEDINGS{Jaspreet2015,
  author={Bhatia, Jaspreet and Breaux, Travis D.},
  booktitle={2015 IEEE Eighth International Workshop on Requirements Engineering and Law (RELAW)}, 
  title={Towards an information type lexicon for privacy policies}, 
  year={2015},
  volume={},
  number={},
  pages={19-24},
  doi={10.1109/RELAW.2015.7330207}}

@inproceedings{zhu2023loghub,
  title = {Loghub: A Large Collection of System Log Datasets for AI-driven Log Analytics},
  author = {Zhu, Jieming and He, Shilin and He, Pinjia and Liu, Jinyang and Lyu, Michael R.},
  booktitle = {IEEE International Symposium on Software Reliability Engineering (ISSRE)},
  year = {2023},
  publisher = {IEEE}
}

@inproceedings{jiang2024loghub2,
  title = {A Large-scale Evaluation for Log Parsing Techniques: How Far are We?},
  author = {Jiang, Zhihan and Liu, Jinyang and Huang, Junjie and Li, Yichen and Huo, Yintong and Gu, Jiazhen and Chen, Zhuangbin and Zhu, Jieming and Lyu, Michael R.},
  booktitle = {ACM SIGSOFT International Symposium on Software Testing and Analysis (ISSTA)},
  year = {2024},
  publisher = {ACM}
}

@article{zimmeck2019maps,
  title={Maps: Scaling privacy compliance analysis to a million apps},
  author={Zimmeck, Sebastian and Story, Peter and Smullen, Daniel and Ravichander, Abhilasha and Wang, Ziqi and Reidenberg, Joel and Russell, N Cameron and Sadeh, Norman},
  journal={Proceedings on privacy enhancing technologies},
  year={2019}
}

@article{ravichander2019question,
  title={Question answering for privacy policies: Combining computational and legal perspectives},
  author={Ravichander, Abhilasha and Black, Alan W and Wilson, Shomir and Norton, Thomas and Sadeh, Norman},
  journal={arXiv preprint arXiv:1911.00841},
  year={2019}
}

@inproceedings{andow2019policylint,
  title={$\{$PolicyLint$\}$: investigating internal privacy policy contradictions on google play},
  author={Andow, Benjamin and Mahmud, Samin Yaseer and Wang, Wenyu and Whitaker, Justin and Enck, William and Reaves, Bradley and Singh, Kapil and Xie, Tao},
  booktitle={28th USENIX security symposium (USENIX security 19)},
  pages={585--602},
  year={2019}
}

@INPROCEEDINGS{Zishuo2023,
  author={Ding, Zishuo and Tang, Yiming and Li, Yang and Li, Heng and Shang, Weiyi},
  booktitle={2023 IEEE/ACM 45th International Conference on Software Engineering (ICSE)}, 
  title={On the Temporal Relations between Logging and Code}, 
  year={2023},
  volume={},
  number={},
  pages={843-854},
  doi={10.1109/ICSE48619.2023.00079}}

@article{Wenhao2017,
title = {DroidInjector: A process injection-based dynamic tracking system for runtime behaviors of Android applications},
journal = {Computers \& Security},
volume = {70},
pages = {224-237},
year = {2017},
issn = {0167-4048},
doi = {https://doi.org/10.1016/j.cose.2017.06.001},
url = {https://www.sciencedirect.com/science/article/pii/S0167404817301207},
author = {Wenhao Fan and Yaohui Sang and Daishuai Zhang and Ran Sun and Yuan'an Liu},
}

@inproceedings{Chen2017,
  author       = {Boyuan Chen and
                  Zhen Ming (Jack) Jiang},
  editor       = {Sebasti{\'{a}}n Uchitel and
                  Alessandro Orso and
                  Martin P. Robillard},
  title        = {Characterizing and detecting anti-patterns in the logging code},
  booktitle    = {Proceedings of the 39th International Conference on Software Engineering,
                  {ICSE} 2017, Buenos Aires, Argentina, May 20-28, 2017},
  pages        = {71--81},
  publisher    = {{IEEE} / {ACM}},
  year         = {2017},
  url          = {https://doi.org/10.1109/ICSE.2017.15},
  doi          = {10.1109/ICSE.2017.15},
}

@inproceedings{Zhiyuan2024Vulnerability,
author = {Chen, Zhiyuan},
title = {A Comprehensive Study of Privacy Leakage Vulnerability in Android App Logs},
year = {2024},
isbn = {9798400712487},
publisher = {Association for Computing Machinery},
address = {New York, NY, USA},
url = {https://doi-org.ezproxy.rit.edu/10.1145/3691620.3695609},
doi = {10.1145/3691620.3695609},
booktitle = {Proceedings of the 39th IEEE/ACM International Conference on Automated Software Engineering},
pages = {2510–2513},
numpages = {4},
location = {Sacramento, CA, USA},
series = {ASE '24}
}

@inproceedings{Stephan2019,
  author       = {Stephan A. Fahrenkrog{-}Petersen and
                  Han van der Aa and
                  Matthias Weidlich},
  title        = {{PRETSA:} Event Log Sanitization for Privacy-aware Process Discovery},
  booktitle    = {International Conference on Process Mining, {ICPM} 2019, Aachen, Germany,
                  June 24-26, 2019},
  pages        = {1--8},
  publisher    = {{IEEE}},
  year         = {2019},
  url          = {https://doi.org/10.1109/ICPM.2019.00012},
  doi          = {10.1109/ICPM.2019.00012},
  bibsource    = {dblp computer science bibliography, https://dblp.org}
}

@online{Josh2024,
	title        = {iPhone vs Android User Stats (2024 Data)},
	author       = {Josh Howarth},
	url          = {https://explodingtopics.com/blog/iphone-android-users},
	urldate      = {2024-03-21},
}

@inproceedings{Christof2016,
  author       = {Christof Rath},
  title        = {Usable Privacy-Aware Logging for Unstructured Log Entries},
  booktitle    = {11th International Conference on Availability, Reliability and Security,
                  {ARES} 2016, Salzburg, Austria, August 31 - September 2, 2016},
  pages        = {272--277},
  publisher    = {{IEEE} Computer Society},
  year         = {2016},
  url          = {https://doi.org/10.1109/ARES.2016.1},
  doi          = {10.1109/ARES.2016.1},
}

@article{west2021best,
  title={Best practice in statistics: Use the Welch t-test when testing the difference between two groups},
  author={West, Robert M},
  journal={Annals of clinical biochemistry},
  volume={58},
  number={4},
  pages={267--269},
  year={2021},
  publisher={SAGE Publications Sage UK: London, England}
}

@ARTICLE{ZhiyuanIEEE,
  author={Chen, Zhiyuan and Nava-Camal, Vanessa and Roy, Tiash and Li, Zhe and Tang, Yiming and Zhang, Xueling and Yang, Haibo},
  journal={IEEE Software}, 
  title={Exploring LLMs’ Potential for Privacy Leakage Detection in Android App Logs: An Empirical Study}, 
  year={2025},
  volume={},
  number={},
  pages={1-6},
  doi={10.1109/MS.2025.3618099}}

\end{document}